\documentstyle[amssymb,11pt]{article}
\begin{document}

\begin{center}{\bf  KINEMATIC QUANTITIES AND RAYCHAUDHURI EQUATIONS IN A $5D$ UNIVERSE} \end{center}
\begin{center}{AUREL BEJANCU \\ Department of Mathematics \\ Kuwait University\\ P.O.Box 5969, Safat 13060\\ Kuwait\\ 
 E-mail:aurel.bejancu@ku.edu.kw}\end{center}

\begin{abstract}
{Based on some ideas emerged from the classical Kaluza-Klein theory, we present a $5D$ universe as a product bundle 
over the $4D$ spacetime. This enables us to introduce and study two categories of kinematic quantities (expansions, 
 shear, vorticity) in a $5D$ universe. One category is related to the fourth dimension (time), and the other one 
comes from the assumption of the existence of the fifth dimension. The Raychaudhuri type equations that we obtain 
in the paper, lead us to results on the evolution of both the $4D$ expansion and $5D$ expansion in a $5D$ universe.}
 \end{abstract}


PACS number(s): 04.50.-h
 
\medskip
\newpage
\section{Introduction}As it well known, the (1+3) threading  of a $4D$ spacetime was developed in order 
to relate physics and geometry to the observations. This theory is based on the existence of a congruence of 
timelike curves and it was successfully applied to: the relativistic cosmology  \cite{ej}, the study of 
 gravito-electromagnetism \cite{b}, the splitting of Einstein equations \cite{ee}, and to some other physical 
theories.\par
In the present paper, we extend the above theory to a (1+1+3) threading of a $5D$ universe. As far as we know, 
there are two important five-dimensional gravity theories: the brane-world gravity \cite{mk}, and the 
space-time-matter theory \cite{o}. In the theory we develop in the present paper, the $4D$ spacetime $M$ is the 
base manifold of a submersion on the $5D$ universe $\bar{M}$, while in the above mentioned theories, $M$ is 
an embedded submanifold of $\bar{M}$. The new geometric configuration of $\bar{M}$ enables us to consider two 
 orthogonal line bundles: the temporal distribution ${\cal{T}}\bar{M}$ which defines a congruence of timelike 
 curves, and the vertical distribution ${\cal{V}}\bar{M}$ which is tangent to a congruence of spacelike curves.
  The spatial distribution ${\cal{S}}\bar{M}$ is the complementary orthogonal distribution to 
${\cal{T}}\bar{M}\oplus {\cal{V}}\bar{M}$ in the tangent bundle of $\bar{M}$. The kinematic quantities (expansion, 
shear, vorticity) in $\bar{M}$, are introduced as spatial tensor fields on $\bar{M}$, which roughly speaking, 
behave like tensor fields on a three-dimensional manifold. 
In order to define a covariant derivative of such tensor fields with respect to any vector field on $\bar{M}$, 
we introduce the Riemannian spatial connection, which is a metric linear connection on ${\cal{S}}\bar{M}$. Finally, 
we obtain three Raychaudhuri type equations which enable us to study the evolution of the expansion in the 
$5D$ universe $\bar{M}$.\par 
Now, we outline the content of the paper. In Section 2 we present the geometric structure of the $5D$ universe 
$(\bar{M}, \bar{g})$  (see (2.12)), and construct the adapted frame and coframe fields $\{\delta/\delta x^0, 
 \delta/\delta x^{\alpha}, \partial/\partial x^4\}$ and $\{\delta x^0, dx^{\alpha}, \delta x^4\},$ respectively. 
Then, in Section 3 we consider the $4D$ velocity $\xi = \delta/\delta x^0$ and the $5D$ velocity ${\eta 
= \partial /\partial x^4}$, and obtain the expression of the line element of $\bar{g}$ with respect to the 
adapted coframe field (cf.(3.4)). The $4D$ and $5D$ kinematic quantities: expansion, shear and vorticity tensor 
fields are defined in Section 4. They are geometric objects on $\bar{M}$, whose local components with respect 
to adapted frame and coframe fields behave as tensor fields on a three-dimensional manifold. The Riemannian 
spatial connection, which we introduce in Section 5, has an important role into the study. Here, we show that 
the Levi-Civita connection of $(\bar{M}, \bar{g})$ is completely determined by the kinematic quantities, 
the Riemannian spatial connection, and the spatial tensor fields we introduced in Section 4 (cf. (5.12)). In 
Section 6 we show that each spatial tensor field defines a tensor field of the same type on $\bar{M}$, and 
evaluate the covariant derivatives of both velocities (cf. (6.9) and (6.10)). In Section 7 we derive the Raychaudhuri 
equations with respect to the expansions $\Theta$ and $K$ of the $5D$ universe (cf. (7.14), (7.16), (7.22)). In 
spite of the full generality approached in the paper, we can see that due to the spatial tensor fields and to the
 Riemannian spatial connection, the calculations can be easily 
handled. By using Raychaudhuri equations we state two results on the evolution of $\Theta$ and $K$ in 
$(\bar{M}, \bar{g})$ (cf. Theorems 7.1, 7.2). Conclusions on the theory we develop in the paper are presented 
in Section 8.
\newpage

\section{Adapted Frames and Coframes in a $5D$ Universe}

Let $M$ and $K$ be manifolds of dimension four and one respectively, and $\bar{M} = M\times K$ be the product 
bundle over $M$ with fibre $K$. Then, a coordinate system $(x^i)$ on $M$ determines a coordinate system $(x^a)= 
(x^i, x^4)$ on $\bar{M}$, where $x^4$ is the fibre coordinate. A general Kaluza-Klein theory on $\bar{M}$ is 
developed with respect to the gauge group $U(1)$, and the coordonate transformations on $\bar{M}$ are given by 

$$(a) \ \ \ \widetilde{x}^{i} = \widetilde{x}^{i}(x^0, x^1, x^2, x^3); \ \ \ (b) \ \ \ \widetilde{x}^4 = x^4 + 
\bar{f}(x^0, x^1, x^2, x^3).\eqno(2.1)$$
As a consequence of (2.1) we deduce that the natural frame fields ${\partial / \partial x^a}$ and ${\partial 
/ \partial \widetilde{x}^a}$ are related by 

$$ (a) \ \ \frac{\partial }{\partial x^{i}} = \frac{\partial \widetilde{x}^{k}}{\partial x^{i}}
\frac{\partial }{\partial \widetilde{x}^{k}} + \frac{\partial \bar{f}}{\partial x^{i}}\frac{\partial }
{\partial \widetilde{x}^{4}}, \ \ \ \ (b)  \ \ \ \frac{\partial }{\partial x^{4}} = \frac{\partial }
{\partial \widetilde{x}^{4}}.\eqno(2.2)$$
Throughout the paper we use the ranges of indices: $a, b, c, ... \in\{0, 1,2,3,4\},$ $i, j, k, ... \in\{0, 1,2,3\},$ 
and $\alpha, \beta, \gamma, ... \in\{1, 2, 3\}$. Also, for any vector bundle $E$ over $\bar{M}$ denote by 
$\Gamma(E)$ the ${\cal{F}}(\bar{M})$-module of smooth sections of $E$, where ${\cal{F}}(\bar{M})$ is the algebra 
of smooth functions on $\bar{M}$.\par
From (2.2b) we see that there exists a globally defined vector field $\eta$ on $\bar{M}$, which is given locally 
by ${\partial / \partial x^4}$. Then denote by ${\cal{V}}\bar{M}$ the line bundle over $\bar{M}$ spanned by 
$\eta$, and call it the {\it vertical distribution} on $\bar{M}$. Suppose that $\bar{M}$ is endowed with a 
Lorentz metric $\bar{g}$ such that

$$\bar{g}(\frac{\partial}{\partial x^4}, \frac{\partial}{\partial x^4}) = \Psi^2, \eqno(2.3) $$
where $\Psi$ is a non-zero function that is globally defined on $\bar{M}$. Denote by ${\cal{H}}\bar{M}$ 
the complementary orthogonal vector bundle to ${\cal{V}}\bar{M}$ in the tangent bundle $T\bar{M}$ of $\bar{M}$, 
and call it the {\it horizontal distribution} on $\bar{M}$. Hence we have the Whitney decomposition 

 $$ T\bar{M} = {\cal{H}}\bar{M}\oplus {\cal{V}}\bar{M}.\eqno(2.4)$$
Now, suppose that on $M$ there exists a globally defined vector field $U$, which induces a special coordinate 
system $(x^i)$ on $M$ such that $U = \partial/\partial x^0.$ Two such coordinate systems 
$(x^i)$ and $(\widetilde{x}^i)$ on $M$  are related by  
 
$$(a) \ \ \ \widetilde{x}^{\alpha} = \widetilde{x}^{\alpha}(x^1, x^2, x^3); \ \ \ (b) \ \ \ \widetilde{x}^0 = x^0 + 
f(x^1, x^2, x^3).\eqno(2.5)$$
Thus, the coordinate transformations on $\bar{M}$ are given by (2.5) and (2.1b). Then, (2.2a) becomes  

$$\begin{array}{l} (a) \ \ \frac{\partial }{\partial x^{\alpha}} = \frac{\partial \widetilde{x}^{\gamma}}
{\partial x^{\alpha}}\frac{\partial }{\partial \widetilde{x}^{\gamma}} + \frac{\partial f}
{\partial x^{\alpha}}\frac{\partial }{\partial \widetilde{x}^0} + \frac{\partial \bar{f}}
{\partial x^{\alpha}}\frac{\partial }{\partial \widetilde{x}^4},\vspace{2mm} \\ (b)  \ \ \ \frac{\partial }
{\partial x^0} = \frac{\partial }{\partial \widetilde{x}^0} + \frac{\partial \bar{f}}{\partial x^0}
\frac{\partial }{\partial \widetilde{x}^4},\end{array}\eqno(2.6)$$
where $\{\partial /\partial x^i\}$ is the lift of the natural frame filed on $M$ to $\bar{M}$.\par
Next, suppose that the lift of $\partial /\partial x^0$ to $\bar{M}$ is timelike with respect to $\bar{g}$ 
and denote by $\delta /\delta{x}^0$ its projection on ${\cal{H}}\bar{M}$ with respect to (2.4). Thus there 
exists locally on $\bar{M}$ a unique function $A_0$, such that 

$$\frac{\delta }{\delta x^0} = \frac{\partial }{\partial x^0} - 
A_0\frac{\partial }{\partial x^4}. \eqno(2.7)$$
By direct calculations, using (2.7), (2.6b) and (2.2b), we deduce that 

$$\frac{\delta }{\delta x^0} = \frac{\delta }{\delta \tilde{x}^0} + (\tilde{A}_0 + \frac{\partial\bar{f}}
{\partial x^0} - A_0)\frac{\partial }{\partial \tilde{x}^4}. $$
Hence

$$(a) \ \ \ \frac{\delta }{\delta x^0} = \frac{\delta }{\delta \tilde{x}^0}, \ \ \ (b) \ \ \ A_0 = \tilde{A}_0 
+ \frac{\partial\bar{f}}{\partial x^0}, \eqno(2.8)$$
with respect to the coordinate transformations on $\bar{M}$. From (2.8a) we conclude that there exists a 
globally defined  horizontal vector field $\xi$ on $\bar{M}$, which is locally given by $\delta/\delta x^0$. 
Moreover, $\xi$ is a timelike vector field on $\bar{M}$. To show this, we consider the line element of $\bar{g}$ 
given by 

$$(a) \ \ \ d\bar{s}^2 = \bar{g}_{ab}dx^adx^b,\ \ \ (b) \ \ \ \bar{g}_{ab} = g\left(\frac{\partial }{\partial x^a}, 
 \frac{\partial }{\partial x^b}\right).\eqno(2.9)$$
Then, taking into account that $\xi$ is orthogonal to $\eta$, and using (2.7), (2.9b) and (2.3), we obtain 

$$A_0 = \Psi^{-2}\bar{g}_{04}.\eqno(2.10)$$
By similar calculations we infer that

$$\bar{g}(\frac{\delta}{\delta x^0}, \frac{\delta}{\delta x^0}) = \bar{g}_{00} - \left(A_0\Psi\right)^2. $$
As $\bar{g}_{00} < 0$, there exists a globally defined non-zero function $\Phi$ on $\bar{M}$ such that 

$$\bar{g}(\frac{\delta}{\delta x^0}, \frac{\delta}{\delta x^0}) = -\Phi^2, \eqno(2.11)$$
and therefore $\xi$ is timelike. Denote by ${\cal{T}}\bar{M}$ the line bundle spanned by $\xi$ and call it the 
{\it temporal distribution} on $\bar{M}$. Thus the decomposition from (2.4) admits the rafinement 

$$ T\bar{M} = {\cal{T}}\bar{M}\oplus {\cal{S}}\bar{M}\oplus {\cal{V}}\bar{M},\eqno(2.12)$$
where ${\cal{S}}\bar{M}$ is the complementary orthogonal bundle to ${\cal{T}}\bar{M}$ in ${\cal{H}}\bar{M}$. 
As $\bar{g}$ is a Lorentz metric, and ${\cal{T}}\bar{M}$ is a timelike vector bundle, we conclude that 
${\cal{S}}\bar{M}$ is spacelike. Thus, we are entitled to call ${\cal{S}}\bar{M}$ the {\it spatial distribution} 
on $\bar{M}$. \par
Next, consider $\{\partial /\partial x^{\alpha}\}, \alpha \in\{1,2,3\}$ , as vector fields locally defined on 
$\bar{M}$ and denote by $\{\delta /\delta x^{\alpha}\}$ their projections on  ${\cal{S}}\bar{M}$ with respect to 
the decomposition (2.12). Thus, we have

$$\frac{\delta }{\delta x^{\alpha}} = \frac{\partial }{\partial x^{\alpha}} - B_{\alpha}\frac{\delta }{\delta x^0} 
- A_{\alpha}\frac{\partial }{\partial x^4}, \eqno(2.13)$$ 
where $A_{\alpha}$ and $B_{\alpha}$ are locally defined functions on $\bar{M}$. By direct calculations, using 
(2.6a), (2.8a), (2.2b) and (2.7) into (2.13), we deduce that 

$$ \begin{array}{l} \frac{\delta }{\delta x^{\alpha}} = \frac{\partial \tilde{x}^{\gamma}}{\partial 
 x^{\alpha}}\frac{\delta }{\delta \tilde{x}^{\gamma}} + \left(\frac{\partial \tilde{x}^{\gamma}}{\partial 
 x^{\alpha}}\tilde{B}_{\gamma} + \frac{\partial f}{\partial x^{\alpha}} - B_{\alpha}\right)
\frac{\delta  } {\delta\tilde{x}^0}\vspace{2mm} \\ \hspace*{10mm}+ \left(\frac{\partial \tilde{x}^{\gamma}}
{\partial x^{\alpha}}\tilde{A}_{\gamma} + \frac{\partial f}{\partial x^{\alpha}}\tilde{A}_0 + 
\frac{\partial \tilde{f}}{\partial x^{\alpha}}- A_{\alpha}\right)\frac{\partial } {\partial\tilde{x}^4},\end{array}$$
 which via (2.12) implies
 
$$ \begin{array}{l}(a) \ \ \  \frac{\delta }{\delta x^{\alpha}} = \frac{\partial \tilde{x}^{\gamma}}{\partial 
 x^{\alpha}}\frac{\delta }{\delta \tilde{x}^{\gamma}}, \ \ \ (b) \ \ \ B_{\alpha} = 
\frac{\partial \tilde{x}^{\gamma}}{\partial  x^{\alpha}}\tilde{B}_{\gamma} + \frac{\partial f}
{\partial x^{\alpha}},\vspace{2mm} \\ \hspace*{10mm}\ \ \ (c) \ \ \ A_{\alpha} = \frac{\partial \tilde{x}^{\gamma}}
{\partial x^{\alpha}}\tilde{A}_{\gamma} + \frac{\partial f}{\partial x^{\alpha}}\tilde{A}_0 + 
\frac{\partial \tilde{f}}{\partial x^{\alpha}}.\end{array}\eqno(2.14)$$ 
By the above geometric construction, we introduce into the study the orthogonal frame field $\{\delta/\delta x^0, 
\delta/\delta x^{\alpha}, \partial/\partial x^4 \}$, which we call {\it adapted frame field} on $\bar{M}$. Its 
dual frame field $\{\delta x^0, dx^{\alpha}, \delta x^4 \}$, where we put

$$(a) \ \ \ \delta x^0 = dx^0 + B_{\alpha}dx^{\alpha},  \ \ \ (b) \ \ \ \delta x^4 = dx^4 + A_idx^i, \eqno(2.15)$$
is called an {\it adapted coframe field} on $\bar{M}$. \par
The pair $(\bar{M}, \bar{g})$ with the geometric configuration described in this section is called a $5D$ 
{\it universe}, and is going to be the main object studied in the present paper. It is important to note that the 
$5D$ universe that we introduce in this paper is different from the ones considered in the well known theories: 
 brane-world theory \cite{mk} and space-time-matter theory \cite{o}. This is because in the present theory the 
$4D$ spacetime is the base manifold of a submersion defined on the $5D$ universe, while in the above theories the 
$4D$ spacetime is considered embedded in the $5D$ universe. 

\section{ $4D$ and $5D$ Velocities in a $5D$ Universe}

Let $(\bar{M}, \bar{g})$ be a $5D$ universe with the line element given by (2.9). From the previous section 
we conclude that $\bar{M}$ admits a double threading by two orthogonal congruences of curves. These congruences 
are defined by the timelike vector field $\xi$ and by the spacelike vector field $\eta$, which we call the $4D$ 
{\it velocity} and $5D$ {\it velocity}, respectively. The name $4D$ velocity for $\xi$ is justified by the fact 
that its integral curves are tangent on their entire length to the horizontal distribution, whose fibres 
are four-dimensional. On the contrary, the integral curves of $\eta$ are orthogonal to ${\cal{H}}\bar{M}$, 
and therefore they are intimately related to the fifth dimension. Moreover, by using (2.7), we obtain 

$$(a) \ \ \left [\frac{\delta }{\delta x^0}, \frac{\partial }{\partial x^4}\right] = 
a_0\frac{\partial }{\partial x^4}, \ \ \ (b) \ \ \ a_0 = \frac{\partial A_0}{\partial x^4}. \eqno(3.1)$$
Thus, the distribution ${\cal{T}}\bar{M}\oplus{\cal{V}}\bar{M}$ is an integrable distribution, and therefore, 
the $5D$ universe admits also a foliation by surfaces, whose transversal bundle (\cite{bf}, p.7) is the 
spatial distribution ${\cal{S}}\bar{M}$.
Next, denote by $h$ the Riemannian metric induced by $\bar{g}$ on ${\cal{S}}\bar{M}$, and put 

$$h_{\alpha\beta} = h\left(\frac{\delta}{\delta x^{\beta}}, \frac{\delta}{\delta x^{\alpha}}\right) 
= \bar{g}\left(\frac{\delta}{\delta x^{\beta}}, \frac{\delta}{\delta x^{\alpha}}\right), \ 
\alpha, \beta\in\{1,2,3\}.\eqno(3.2)$$
Then, by using (2.9b), (2.13), (3.2), (2.11) and (2.3), we deduce that 

$$h_{\alpha\beta} = \bar{g}_{\alpha\beta} + \Phi^2B_{\alpha}B_{\beta} - \Psi^2A_{\alpha}A_{\beta}.\eqno(3.3) $$
Due to (2.3), (2.11) and (3.3), the line element with respect to the adapted coframe field has  the simple form 

$$d\bar{s}^2 = -\Phi^2(\delta x^0)^2 + h_{\alpha\beta}dx^{\alpha}dx^{\beta} + \Psi^2(\delta x^4)^2.\eqno(3.4) $$
Now, in order to find the covariant local components of the above velocities we consider the 1-forms $\xi^{\star}$ 
and $\eta^{\star}$ given by 

$$(a) \ \ \ \xi^{\star}(X) = \bar{g}(X, \xi), \ \ \ (b) \ \ \ \eta^{\star}(X) = \bar{g}(X, \eta), \ \ \forall 
\ X \in \Gamma(T\bar{M}). \eqno(3.5) $$
Then we put 

$$(a) \ \ \ \xi_a = \xi^{\star}(\frac{\partial }{\partial x^a}), \ \ \ (b) \ \ \ \eta_a = \eta^{\star}(\frac{
\partial }{\partial x^a}), \ \ a \in \{0,1,2,3,4\}, \eqno(3.6)$$
and by using (3.6), (3.5), (2.7), (2.11) , (2.13) and (2.3), we obtain 

$$(a) \ \ \ \xi_0 = -\Phi^2, \ \ \ (b) \ \ \ \xi_{\alpha} = -\Phi^2B_{\alpha}, \ \ \ (c) \ \ \ \xi_4 = 0, \ 
\forall \alpha \in \{1,2,3\}, \eqno(3.7) $$
and 

$$(a) \ \ \ \eta_i = \Psi^2A_i, \ \ \ (b) \ \ \ \eta_4 = \Psi^2, \ \forall \ i \in \{0,1,2,3,\}.\eqno(3.8) $$
By using (3.8a) and (3.7b) into (2.7) and (2.13), we infer that 

$$\begin{array}{c} (a) \ \ \ \frac{\delta }{\delta x^0} = \frac{\partial }{\partial x^0} - 
\Psi^{-2}\eta_0\frac{\partial }{\partial x^4}, \vspace{2mm} \\ (b) \ \ \ \frac{\delta }{\delta x^{\alpha}} 
= \frac{\partial }{\partial x^{\alpha}} + \Phi^{-2}\xi_{\alpha}\frac{\delta }{\delta x^0} 
- \Psi^{-2}\eta_{\alpha}\frac{\partial }{\partial x^4}.\end{array} \eqno(3.9)$$ 
Similarly, (2.15) becomes 

$$(a) \ \ \ \delta x^0 = dx^0 - \Phi^{-2}\xi_{\alpha}dx^{\alpha},  \ \ \ (b) \ \ \ \delta x^4 = dx^4 
+ \Psi^{-2}\eta_idx^i. \eqno(3.10)$$
In what follows, we also call $\xi^{\star} = (\xi_a)$ and $\eta^{\star} = (\eta_a)$ the $4D$ {\it velocity} and 
$5D$ {\it velocity} in $(\bar{M}, \bar{g})$, respectively.

\section{Spatial Tensor Fields and Kinematic Quantities in a $5D$ Universe}

In this section we introduce spatial tensor fields in $(\bar{M}, \bar{g})$ as geometric objects whose local 
components behave as the ones of tensor fields on a three-dimensional manifold. In particular, we define 
the expansion, shear and vorticity tensor fields as spatial tensor fields.\par
First, by using (2.14a) into (3.2), we deduce that the local components of the Riemannian metric $h$ 
on ${\cal{S}}\bar{M}$ satisfy

$$h_{\alpha\beta} = \widetilde{h}_{\mu\nu}\frac{\partial\tilde{x}^{\mu}}{\partial x^{\alpha}}
\frac{\partial\tilde{x}^{\nu}}{\partial x^{\beta}}, \ \ \alpha, \beta \in \{1,2,3\}, \eqno(4.1)$$
with respect to the coordinate transformations on $\bar{M}$. Also, the entries of the inverse of the $3\times 3$ 
matrix $[h_{\alpha\beta}]$ satisfy

$$\widetilde{h}^{\mu\nu} = h^{\alpha\beta}\frac{\partial\tilde{x}^{\mu}}{\partial x^{\alpha}}
\frac{\partial\tilde{x}^{\nu}}{\partial x^{\beta}}, \ \ \mu, \nu \in \{1,2,3\}. \eqno(4.2)$$
Thus, $h_{\alpha\beta}$ and $h^{\alpha\beta}$ are locally functions on the five-dimensional manifold $\bar{M}$, 
but they are transformed as the local components of some tensor fields of type $(0, 2)$ and $(2, 0)$ on a 
 three-dimensional manifold. This leads us to an important category of geometric objects on $\bar{M}$. Namely, 
we say that the functions $T^{\gamma_1 \cdots \gamma_p}_{\alpha_1\cdots \alpha_q}(x^a)$ define a spatial tensor 
field of type $(p, q)$ in the $5D$ universe $(\bar{M}, \bar{g})$, if they satisfy 

$$T_{\alpha_1 \cdots \alpha_q}^{\gamma_1\cdots \gamma_p}\frac{\partial\tilde{x}^{\mu_1}}{\partial x^{\gamma_1}}
\cdots \frac{\partial\tilde{x}^{\mu_p}}{\partial x^{\gamma_p}}
= \tilde{T}^{\mu_1 \cdots \mu_p}_{\nu_1\cdots \nu_q}\frac{\partial\tilde{x}^{\nu_1}}{\partial x^{\alpha_1}}
\cdots \frac{\partial\tilde{x}^{\nu_q}}{\partial x^{\alpha_q}},\eqno(4.3)$$
with respect to the coordinate  transformations on $\bar{M}$. By using (4.1) and (4.2), we see that  
$h_{\alpha\beta}$ and $h^{\alpha\beta}$ define spatial tensor fields of types (0,2) and (2,0), respectively.\par
Next, by direct calculations using (2.7) and (2.13), we obtain 

$$\begin{array}{lc}(a) \ \ \left[\frac{\delta }{\delta x^{\alpha}}, \frac{\delta }{\delta x^0}\right] 
= b_{\alpha}\frac{\delta }{\delta x^0} + a_{\alpha}\frac{\partial }{\partial x^4},\vspace{2mm}\\ (b) \ \ 
 \left[\frac{\delta }{\delta x^{\alpha}}, \frac{\partial }{\partial x^4}\right] = 
d_{\alpha}\frac{\delta }{\delta x^0} + c_{\alpha}\frac{\partial }{\partial x^4}, \vspace{2mm} \\ (c) \ \ \ 
\left[\frac{\delta }{\delta x^{\beta}} ,\frac{\delta }{\delta x^{\alpha}}\right] = 2\omega_{\alpha\beta}
\frac{\delta }{\delta x^0} + 2\eta_{\alpha\beta}\frac{\partial }{\partial x^4}, \end{array}\eqno(4.4)$$
where we put 

$$\begin{array}{lc} 
(a) \ \ \ a_{\alpha} = \frac{\delta A_{\alpha}}{\delta x^0} - \frac{\delta A_0}{\delta x^{\alpha}} - 
B_{\alpha}\frac{\delta A_0}{\delta x^0}, \ \ \ (b) \ \ \ b_{\alpha} = \frac{\delta B_{\alpha}}{\delta x^0}, 
 \vspace{2mm}\\ (c) \ \ \ c_{\alpha} = \frac{\partial A_{\alpha}}{\partial x^4} -  
B_{\alpha}\frac{\partial A_0}{\partial x^4}, \ \ \ (d) \ \ \ d_{\alpha} = \frac{\partial B_{\alpha}}{\partial x^4}, 
 \vspace{2mm}\\
(e) \ \ \ \omega_{\alpha\beta} = \frac{1}{2}\left\{\frac{\delta B_{\beta}}{\delta x^{\alpha}} - 
\frac{\delta B_{\alpha}}{\delta x^{\beta}}\right\}, \vspace{2mm}\\ 
(f)\ \ \ \eta_{\alpha\beta} = \frac{1}{2}\left\{\frac{\delta A_{\beta}}{\delta x^{\alpha}} - 
\frac{\delta A_{\alpha}}{\delta x^{\beta}} + B_{\alpha}\frac{\delta A_0}{\delta x^{\beta}} - B_{\beta}
\frac{\delta A_0}{\delta x^{\alpha}}\right\}. \end{array}\eqno(4.5)$$
By using (2.2b), (2.8a) and (2.14a) into (4.4), it is easy to check that $a_{\alpha}, b_{\alpha}, 
c_{\alpha}, d_{\alpha}$ define spatial tensor fields of type (0, 1), while $\omega_{\alpha\beta}$ and 
 $\eta_{\alpha\beta}$ define skew-symmetric spatial tensor fields of type (0, 2). Moreover, from (4.4c) we deduce 
that {\it the spatial distribution is integrable, if and only if}, 

$$\omega_{\alpha\beta} = 0, \ \ \ \mbox{and} \ \ \ \eta_{\alpha\beta} = 0, \ \forall \ \alpha, \beta \in \{1,2,3\}.
\eqno(4.6)$$
Thus, extending the terminology from (1+3) threading of a $4D$ spacetime (\cite{ej}, p.81), we call 
 $\omega_{\alpha\beta}$ and $\eta_{\alpha\beta}$ the $4D$ {\it vorticity tensor field} and the $5D$ {\it 
vorticity tensor field}, respectively. The prefix $4D$ is placed in front of vorticity to emphasize that this 
object comes from the structure of the $4D$ spacetime $M$. On the contrary, the $5D$ vorticity is determined by 
the existence of the fifth dimension. Throughout the paper, we use this rule for some other geometric objects.\par
Now, we define the spatial tensor fields of type (0,1):

$$(a) \ \ \ \phi_{\alpha} = \Phi^{-1}\frac{\delta\Phi}{\delta x^{\alpha}}, \ \ \ (b) \ \ \ 
\psi_{\alpha} = \Psi^{-1}\frac{\delta\Psi}{\delta x^{\alpha}},\eqno(4.7)$$
and by using (3.7b) and (3.8a) into (4.5e) and (4.5f), we deduce that

$$\begin{array}{lc} 
(a) \ \ \ \omega_{\alpha\beta} 
= \Phi^{-2}\left\{\phi_{\alpha}\xi_{\beta} - \phi_{\beta}\xi_{\alpha} + \frac{1}{2}\left(\frac{\delta\xi_{\alpha}}{\delta x^{\beta}} - \frac{\delta\xi_{\beta}}{\delta x^{\alpha}}
\right)\right\},\vspace{2mm}\\
(b) \ \ \ \eta_{\alpha\beta} 
= \Psi^{-2}\left\{\eta_{\alpha}\psi_{\beta} - \eta_{\beta}\psi_{\alpha} + 
 \frac{1}{2}\left(\frac{\delta\eta_{\beta}}{\delta x^{\alpha}} - \frac{\delta\eta_{\alpha}}{\delta x^{\beta}}
\right)\right.  \\ \left. \hspace*{18mm} + \ \Phi^{-2}\eta_0\left(\xi_{\alpha}\psi_{\beta} - 
 \xi_{\beta}\psi_{\alpha}\right) + \frac{1}{2}\Phi^{-2}\left(\xi_{\beta}\frac{\delta\eta_0}{\delta x^{\alpha}} 
- \xi_{\alpha}\frac{\delta\eta_0}{\delta x^{\beta}}\right)\right\}.  \end{array}\eqno(4.8)$$
 Next, we denote by ${\cal{L}}$ the Lie derivative on $\bar{M}$ and define the functions 
 
 $$\begin{array}{c}(a)\ \ \ \Theta_{\alpha\beta} = \frac{1}{2}\left({\cal{L}}_{
 \frac{\delta}{\delta x^0}}\bar{g}\right)\left(\frac{\delta }{\delta x^{\beta}},  
\frac{\delta }{\delta x^{\alpha}}\right),  \vspace{3mm} \\ 
(b) \ \ \ K_{\alpha\beta} = \frac{1}{2}\left({\cal{L}}_{
 \frac{\partial  }{\partial x^4}}\bar{g}\right)\left(\frac{\delta }{\delta x^{\beta}},  
\frac{\delta }{\delta x^{\alpha}}\right).\end{array} \eqno(4.9)$$
 Then, by using (3.2), (4.4a) and (4.4b) into (4.9), we obtain
 
 $$(a) \ \ \Theta_{\alpha\beta} = \frac{1}{2}\frac{\delta h_{\alpha\beta}}{\delta x^0}, \ \ \ (b) \ \ \ 
K_{\alpha\beta} = \frac{1}{2}\frac{\partial h_{\alpha\beta}}{\partial x^4}. \eqno(4.10)$$
Moreover, applying $\delta/\delta x^0$ and $\partial/\partial x^4$ to (4.1), and taking into account (2.5a), 
(2.8a) and (2.2b), we deduce that $\Theta_{\alpha\beta}$ and $K_{\alpha\beta}$ define symmetric spatial tensor 
fields of type (0,2). We call $\Theta_{\alpha\beta}$ and $K_{\alpha\beta}$ the $4D$ {\it expansion tensor field} 
and $5D$ {\it expansion tensor field}, respectively. Taking the traces of these tensor fields, we obtain the 
$4D$ {\it expansion function} $\Theta$ and the $5D$ {\it expansion function} $K$, given by 

$$(a) \ \ \ \Theta = \Theta_{\alpha\beta}h^{\alpha\beta}, \ \ \ (b) \ \ \ K 
= K_{\alpha\beta}h^{\alpha\beta}.\eqno(4.11) $$
Finally, we define the trace-free symmetric spatial tensor fields

$$(a) \ \ \ \sigma_{\alpha\beta} = \Theta_{\alpha\beta} - \frac{1}{3}\Theta h_{\alpha\beta}, \ \ \ (b) \ \ \ 
H_{\alpha\beta} = K_{\alpha\beta} - \frac{1}{3}K h_{\alpha\beta}.\eqno(4.12) $$
Then, inspired by the terminology from the kinematic theory in a $4D$ spacetime, we call $\sigma_{\alpha\beta}$ 
and $H_{\alpha\beta}$ the $4D$ {\it shear tensor field} and $5D$ {\it shear tensor field}, respectively.\par
As a conclusion of this section, we may say that $\{\omega_{\alpha\beta}, \Theta_{\alpha\beta}, \Theta, 
 \sigma_{\alpha\beta}\}$ and $\{\eta_{\alpha\beta}, K_{\alpha\beta}, K, H_{\alpha\beta}\}$  are the $4D$ {\it 
 kinematic quantities} and the $5D$ {\it kinematic quantities} in the $5D$ universe $(\bar{M}, \bar{g})$, with 
 respect to the congruence of curves defined by $\xi$ and $\eta$, respectively.
 
\section{The Riemannian Spatial Connection in a $5D$ Universe} 

Let $\bar{M}, \bar{g})$ be a $5D$ universe, and ${\cal{S}}\bar{M}^{\star}$ be the dual bundle of the spatial 
 distribution ${\cal{S}}\bar{M}$. Suppose that 
 
$$T:\Gamma({\cal{S}}\bar{M}^{\star})^p \times \Gamma({\cal{S}}\bar{M})^q \longrightarrow {\cal{F}}(\bar{M})
,\eqno(5.1) $$ 
is a $(p+q)-{\cal{F}}(\bar{M})$-multilinear mapping, and locally define the functions 
 
$$T_{\alpha_1 \cdots \alpha_q}^{\gamma_1\cdots \gamma_p} = T(dx^{\gamma_1}, \cdots, dx^{\gamma_p}, 
\frac{\delta }{\delta x^{\alpha_1}}, \cdots, \frac{\delta }{\delta x^{\alpha_q}}).\eqno(5.2)$$ 
 Then, it is easy to check that $T_{\alpha_1 \cdots \alpha_q}^{\gamma_1\cdots \gamma_p}$ satisfy (4.3), and 
therefore they define a spatial tensor field of type $(p, q)$. Conversely, suppose $T_{\alpha_1 \cdots 
 \alpha_q}^{\gamma_1\cdots \gamma_p}$ are functions satisfying (4.3). Then, $T$ defined locally by (5.2) 
becomes  an ${\cal{F}}(\bar{M})$-multilinear mapping as in (5.1).\par
The purpose of this section is to define covariant derivatives of the spatial tensor fields given either as in 
(5.1), or by their local components from (5.2). First, we consider the Levi-Civita connection $\bar{\nabla}$ on 
$(\bar{M}, \bar{g})$ given by (cf. \cite{on}, p.61)

$$\begin{array}{lc}
2\bar{g}(\bar{\nabla}_XY, Z) =  X(\bar{g}(Y, Z)) + Y(\bar{g}(Z, X)) - Z(\bar{g}(X, Y)) \vspace{4mm} 
\\+ \bar{g}([X, Y], Z) -  \bar{g}([Y, Z], X) +   \bar{g}([Z, X], Y),\end{array}\eqno(5.3) $$
for all $X, Y, Z \in \Gamma(T\bar{M})$. Then define the operator 

$$\begin{array}{l}
\nabla:\Gamma(T\bar{M})\times \Gamma({\cal{S}}\bar{M})\longrightarrow \Gamma({\cal{S}}\bar{M}),\vspace{2mm}\\ 
 \nabla(X, sY) = \nabla_XsY = s\bar{\nabla}_XsY, \ \ \ \forall \ \ X,Y \in \Gamma(T\bar{M}),\end{array}\eqno(5.4)$$ 
where $s$ is the projection morphism of $T\bar{M}$ on ${\cal{S}}\bar{M}$ with respect to (2.12). It is easy to 
check that $\nabla$ is a metric linear connection on ${\cal{S}}\bar{M}$, that is we have 

$$(\nabla_Xh)(sY, sZ) = 0, \ \ \ \forall \ X, Y, Z \in \Gamma(T\bar{M}),\eqno(5.5)$$
where $h$ is the Riemannian metric on ${\cal{S}}\bar{M}$. We call $\nabla$ the {\it Riemannian spatial connection} 
in the $5D$ universe $(\bar{M}, \bar{g})$ .\par
Taking into account that all the kinematic quantities have been defined by their local components, we need 
to characterize $\nabla$ by its local coefficients with respect to an adapted frame field. First, we put
 
$$\begin{array}{c}
(a) \ \ \nabla_{\frac{\delta }{\delta x^{\beta}}}\frac{\delta }{\delta x^{\alpha}} = 
\Gamma^{\ \gamma}_{{\alpha}\ \;{\beta}}\frac{\delta }{\delta x^{\gamma}}, \ \ \ (b) \ \ \ \ 
\nabla_{\frac{\partial }{\partial x^0}}\frac{\delta }{\delta x^{\alpha}} =  
\Gamma^{\ \gamma}_{\alpha \ \;0}\frac{\delta }{\delta x^{\gamma}}\vspace{2mm}\\ 
(c) \ \ \nabla_{\frac{\partial }{\partial x^4}}\frac{\delta }{\delta x^{\alpha}} = 
\Gamma^{\ \gamma}_{\alpha \ \;4}\frac{\delta }{\delta x^{\gamma}}.\end{array}\eqno(5.6) $$
Then, take $X = \delta/\delta x^{\beta}, Y = \delta/\delta x^{\alpha}$ and $Z = \delta/\delta x^{\mu}$ in (5.3), and by 
using (5.4), (5.6a), (3.2) and (4.4c), we obtain

$$\Gamma^{\ \gamma}_{\alpha\ \;\beta} = \frac{1}{2}h^{\gamma \mu}\left\{\frac{\delta h_{\mu\alpha}}
{\delta x^{\beta}} +  
\frac{\delta h_{\mu\beta}}{\delta x^{\alpha}} - \frac{\delta h_{\alpha\beta}}{\delta x^{\mu}}\right\}. \eqno(5.7) $$
Note that formally, $\Gamma^{\ \gamma}_{\alpha\ \;\beta}$ look like the Christoffel symbols for a Levi-Civita 
connection on a three-dimensional manifold, but two main differences should be pointed out:\par
(i) In general, $h_{\alpha\beta}$ are functions of all five variables $(x^a)$, \par
(ii) The usual partial derivatives are replaced here by the operators defined by (2.13).\par
Throughout the paper, we use $h_{\alpha\beta}$ and $h^{\alpha\beta}$ for lowering and raising Greek indices. 
As example, for vorticity tensor fields we have 

$$\begin{array}{l}(a) \ \ \ \omega_{\beta}^{\gamma} = h^{\gamma\alpha}\omega_{\alpha\beta}, \ \ \ (b) \ \ \ 
\omega^{\gamma\mu} = h^{\gamma\alpha}h^{\mu\beta}\omega_{\alpha\beta}, \vspace{2mm}\\ 
(c) \ \ \ \eta_{\beta}^{\gamma} = h^{\gamma\alpha}\eta_{\alpha\beta}, \ \ \ (d) \ \ \ 
\eta^{\gamma\mu} = h^{\gamma\alpha}h^{\mu\beta}\eta_{\alpha\beta}.\end{array}\eqno(5.8)$$
Next, we take $Y = \delta/\delta x^{\alpha},$  $Z = \delta/\delta x^{\mu}$, and in turn $X = \delta/\delta x^0,$ 
and $X = \partial/\partial x^4$ in (5.3), and by using (5.4), (5.6b), (5.6c), (3.2), (4.4) and (4.10), we 
deduce that     

$$\begin{array}{lc}(a) \ \ \ 
\Gamma^{\ \gamma}_{\alpha \ \;0} = \Theta_{\alpha}^{\gamma} + \Phi^2\omega_{\alpha}^{\gamma},
\vspace{2mm}\\ \ \ \ (b) \ \ \ \Gamma^{\ \gamma}_{\alpha \ \;4} = K_{\alpha}^{\gamma} 
- \Psi^2\eta_{\alpha}^{\gamma}.\end{array}\eqno(5.9) $$
Now, let $T$ be a spatial tensor field of type $(p,q)$. Then  $\nabla_{\frac{\delta}{\delta x^{\beta}}}T$ 
is a spatial tensor of type $(p, q+1)$, while $\nabla_{\frac{\partial}{\partial x^0}}T$ and 
 $\nabla_{\frac{\partial}{\partial x^4}}T$ are spatial tensor fields 
of the same type $(p, q)$. In particular take   $T = (T_{\alpha}^{\gamma})$ and express these three types of 
covariant derivatives as follows:

$$ \begin{array}{c}(a) \ \ \ 
T_{\alpha\vert_{\beta}}^{\gamma} = \frac{\delta T_{\alpha}^{\gamma}}{\delta x^{\beta}} + T_{\alpha}^{\mu}
\Gamma^{ \  \gamma}_{\mu\ \;\beta} - T_{\mu}^{\gamma}\Gamma^{\;\mu}_{{\alpha}\ \;\beta},\vspace{3mm}\\ (b) \ \ \ 
 T_{{\alpha}\vert_{0}}^{\gamma} = \frac{\delta T_{\alpha}^{\gamma}}{\delta x^0} + T_{\alpha}^{\mu}
\Gamma^{ \ \gamma}_{\mu\ \;0} - T_{\mu}^{\gamma}\Gamma^{\ \mu}_{{\alpha}\ \;0}, 
\vspace{3mm}\\ (c) \ \ \ 
 T_{{\alpha}\vert_{4}}^{\gamma} = \frac{\partial T_{\alpha}^{\gamma}}{\partial x^4} + 
T_{\alpha}^{\mu}\Gamma^{ \ \gamma}_{\mu\ \;4}
  - T_{\mu}^{\gamma}\Gamma^{\ \mu}_{{\alpha}\ \;4}.\end{array}  \eqno(5.10)$$

As $\nabla$ is a metric connection on ${\cal{S}}\bar{M}$, we have:

$$(a) \ \ \ h_{\alpha\beta\vert_{a}} = 0, \ \ \ (b) \ \ \ h^{\alpha\beta}_{\ \ \vert_{a}} 
= 0,\ \forall \alpha, \beta \ \in \{1,2,3\}, \ \ a \ \in \{0,1,2,3,4\}.\eqno(5.11)$$

Finally, by using (5.3), the spatial tensor fields introduced in the previous section, and the local coefficients 
of the Riemannian spatial connection, we express  the Levi-Civita connection of the $5D$ universe $(\bar{M}, \bar{g})$, 
as follows:

$$\begin{array}{lc}
(a) \ \ \bar{\nabla}_{\frac{\delta }{\delta x^{\beta}}}\frac{\delta }{\delta x^{\alpha}} =  
\Gamma^{\ \;\gamma}_{{\alpha}\ \;{\beta}} \frac{\delta }{\delta x^{\gamma}} + \left(\omega_{{\alpha}{\beta}} 
+ \Phi^{-2}\Theta_{{\alpha}{\beta}}\right)\frac{\delta }{\delta x^0} 
\vspace{2mm}\\ \hspace*{26mm} + \left(\eta_{{\alpha}{\beta}} 
- \Psi^{-2}K_{{\alpha}{\beta}}\right)\frac{\partial }{\partial x^4},\vspace{2mm}\\  
(b) \ \ \bar{\nabla}_{\frac{\delta }{\delta x^0}}\frac{\delta }{\delta x^{\alpha}} =  
\Gamma^{\ \;\gamma}_{{\alpha}\ \;0} \frac{\delta }{\delta x^{\gamma}} + \left(\phi_{{\alpha}} - 
b_{\alpha}\right)\frac{\delta }{\delta x^0} \vspace{2mm}\\ \hspace*{26mm} + 
\frac{1}{2}\left(\Phi^2d_{\alpha}\Psi^{-2} - a_{\alpha}\right)\frac{\partial }{\partial x^4},\vspace{2mm}\\ 
(c) \ \ \bar{\nabla}_{\frac{\partial }{\partial x^4}}\frac{\delta }{\delta x^{\alpha}} =  
\Gamma^{\ \;\gamma}_{{\alpha}\ \;4} \frac{\delta }{\delta x^{\gamma}} + 
\frac{1}{2}\left(\Psi^2a_{\alpha}\Phi^{-2} - d_{\alpha}\right)\frac{\delta }{\delta x^0} \vspace{2mm}\\ 
 \hspace*{26mm} + \left(\psi_{{\alpha}} - c_{\alpha}\right)\frac{\partial }{\partial x^4}, \vspace{2mm}\\ 
(d) \ \ \bar{\nabla}_\frac{\delta }{\delta x^{\alpha}}{\frac{\delta }{\delta x^0}} =  
\Gamma^{\ \;\gamma}_{{\alpha}\ \;0}\frac{\delta }{\delta x^{\gamma}} + \phi_{{\alpha}}\frac{\delta }{\delta x^0} + 
\frac{1}{2}\left(\Phi^2d_{\alpha}\Psi^{-2} + a_{\alpha}\right)\frac{\partial }{\partial x^4},\vspace{2mm}\\ 
(e) \ \ \bar{\nabla}_{\frac{\delta }{\delta x^{\alpha}}}{\frac{\partial }{\partial x^4}} =  
\Gamma^{\ \;\gamma}_{{\alpha}\ \;4} \frac{\delta }{\delta x^{\gamma}} + 
\frac{1}{2}\left(\Psi^2a_{\alpha}\Phi^{-2} + d_{\alpha}\right)\frac{\delta }{\delta x^0} 
+ \psi_{{\alpha}}\frac{\partial }{\partial x^4}, \vspace{2mm}\\ 
(f) \ \ \bar{\nabla}_{\frac{\partial }{\partial x^4}}\frac{\delta }{\delta x^0} = \frac{1}{2}\left(\Psi^2a^{\gamma} - 
\Phi^{2}d^{\gamma}\right)\frac{\delta }{\delta x^{\gamma}} + \phi_4\frac{\delta }{\delta x^0} + 
(\psi_0 - a_0)\frac{\partial }{\partial x^4}, \vspace{2mm}\\  
(g) \ \ \bar{\nabla}_\frac{\delta }{\delta x^0}{\frac{\partial }{\partial x^4}} = \frac{1}{2}\left(\Psi^2a^{\gamma} - 
\Phi^{2}d^{\gamma}\right)\frac{\delta }{\delta x^{\gamma}} + \phi_4\frac{\delta }{\delta x^0} + 
\psi_0\frac{\partial }{\partial x^4}, \vspace{2mm}\\
(h) \ \ \bar{\nabla}_\frac{\delta }{\delta x^0}{\frac{\delta }{\delta x^0}} = \Phi^2\left(\phi^{\gamma} - 
b^{\gamma}\right)\frac{\delta }{\delta x^{\gamma}} + \phi_0\frac{\delta }{\delta x^0} + 
\Phi^2\phi_4\Psi^{-2}\frac{\partial }{\partial x^4}, \vspace{2mm}\\
(i) \ \ \bar{\nabla}_\frac{\partial }{\partial x^4}{\frac{\partial }{\partial x^4}} = \Psi^2\left(c^{\gamma} - 
\psi^{\gamma}\right)\frac{\delta }{\delta x^{\gamma}} + \Psi^2(\psi_0 - a_0)\Phi^{-2}\frac{\delta }{\delta x^0} + 
\psi_4\frac{\partial }{\partial x^4}, \end{array}\eqno(5.12)$$
where we put

$$\begin{array}{lc}(a) \ \ \ \phi_0 = \Phi^{-1}\frac{\delta\Phi}{\delta x^0}, \ \ \ (b) \ \ \ \phi_4 
= \Phi^{-1}\frac{\partial\Phi}{\partial x^4}, \vspace{2mm} \\ (c) \ \ \ \psi_0 = \Psi^{-1}\frac{\delta\Psi}{\delta x^0}, 
\ \ \ (d) \ \ \ \psi_4 = \Psi^{-1}\frac{\partial\Psi}{\partial x^4}.\end{array}\eqno(5.13)$$

\section{Covariant Derivatives of $4D$ and $5D$ Velocities}

In this section we show that the covariant derivatives of both the $4D$ velocity $\xi^{\star}$ and $5D$ velocity 
 $\eta^{\star}$ are completely determined by the kinematic quantities and the spatial tensor fields we introduced 
in Sections 4 and 5. Also, we compare the results with what is known in the (1+3) threading of the $4D$ spacetime.\par
First, by using (3.5) and taking into account that $\bar{\nabla}$ is a metric connection, we obtain

$$\begin{array}{l}
(a) \ \ \ 
(\bar{\nabla}_X\xi^{\star})(Y) = \bar{g}(Y, \bar{\nabla}_X\xi),\vspace{2mm}\\ 
(b) \ \ \ 
(\bar{\nabla}_X\eta^{\star})(Y) = \bar{g}(Y, \bar{\nabla}_X\eta),
 \ \ \ \forall \ X, Y \in \Gamma(T\bar{M}). \end{array}\eqno(6.1)$$
 Then, consider the $4D$ {\it acceleration} ${\stackrel{.}{\xi}}_a$ and the $5D$ {\it acceleration}
 ${\stackrel{.}{\eta}}_a$, given by

$$\begin{array}{l}(a) \ \ \ 
{\stackrel{.}{\xi}}_a = \left(\bar{\nabla}_{\frac{\delta}{\delta x^0}}\xi^{\star}\right) 
 \left(\frac{\partial}{\partial x^a}\right), \ \ \  
(b) \ \ \ {\stackrel{.}{\eta}}_a = \left(\bar{\nabla}_{\frac{\partial}{\partial x^4}}\eta^{\star}\right) 
 \left(\frac{\partial}{\partial x^a}\right).\end{array}\eqno(6.2)$$
Next, by using (3.9) we express the natural frame field $\{\partial/\partial x^a\}$ in terms of the adapted 
frame field, as follows:
 
$$\frac{\partial }{\partial x^a} = \delta^{\alpha}_a\frac{\delta }{\delta x^{\alpha}} - \Phi^{-2}\xi_a
\frac{\delta }{\delta x^0} + \Psi^{-2}\eta_a\frac{\partial }{\partial x^4}.\eqno(6.3)$$ 
Then, by using (6.2), (6.1), (6.3), (5.12h) and (5.12i), we infer that  
 
$$\begin{array}{l}(a) \ \ \ {\stackrel{.}{\xi}}_a = \Phi^2\delta_a^{\alpha}(\phi_{\alpha} - b_{\alpha}) + 
\phi_0\xi_a + \phi_4\Phi^2\eta_a\Psi^{-2},\vspace{2mm} \\ 
(b) \ \ \ {\stackrel{.}{\eta}}_a = \Psi^2\delta_a^{\alpha}(c_{\alpha} - \psi_{\alpha}) + \Psi^2(\psi_0 
- a_0)\Phi^{-2}\xi_a + \psi_4\eta_a.\end{array} \eqno(6.4)$$
Now, by using (6.1), (6.3), (5.12) and (5.9), we obtain

$$\begin{array}{l}(a) \ \ \ 
\left(\bar{\nabla}_{\frac{\partial}{\partial x^b}}\xi^{\star}\right)\left(\frac{\delta}{\delta x^{\alpha}}\right) 
= \delta_b^{\gamma}(\Theta_{\alpha\gamma} + \Phi^2\omega_{\alpha\gamma}) + \xi_b(b_{\alpha} - \phi_{\alpha}) 
\vspace{2mm} \\ \hspace*{40mm}+ \frac{1}{2}\eta_b(a_{\alpha} - \Phi^2d_{\alpha}\Psi^{-2}), \vspace{2mm} \\ 
(b) \ \ \ \left(\bar{\nabla}_{\frac{\partial }{\partial x^b}}\xi^{\star}\right)\left(\frac{\delta}{\delta x^0}\right) 
= -\Phi^2\delta_b^{\gamma}\phi_{\gamma} + \phi_0\xi_b - \phi_4\Phi^2\eta_b\Psi^{-2}, \vspace{3mm} \\ 
(c) \ \ \ 
\left(\bar{\nabla}_{\frac{\partial}{\partial x^b}}\xi^{\star}\right)\left(\frac{\partial}{\partial x^4}\right) 
= \frac{1}{2}\delta_b^{\gamma}(\Phi^2d_{\gamma} + \Psi^2a_{\gamma}) - \phi_4\xi_b +  
\vspace{2mm} \\ \hspace*{40mm}+ (\psi_0 - a_0)\eta_b, \end{array}\eqno(6.5)$$
and 
$$\begin{array}{l}(a) \ \ \ 
\left(\bar{\nabla}_{\frac{\partial}{\partial x^b}}\eta^{\star}\right)\left(\frac{\delta}{\delta x^{\alpha}}\right) 
= \delta_b^{\gamma}(K_{\alpha\gamma} - \Psi^2\eta_{\alpha\gamma}) + \frac{1}{2}\xi_b(d_{\alpha} - \Psi^2a_{\alpha}
\Phi^{-2})\vspace{2mm} \\ \hspace*{40mm} + \eta_b(c_{\alpha} -\psi_{\alpha}), 
 \vspace{2mm} \\ 
(b) \ \ \ \left(\bar{\nabla}_{\frac{\partial }{\partial x^b}}\eta^{\star}\right)\left(\frac{\delta}{\delta x^0}\right)
 = -\frac{1}{2}\delta_b^{\gamma}(\Phi^2d_{\gamma} + \Psi^2a_{\gamma}) + \phi_4\xi_b   
\vspace{2mm} \\ \hspace*{40mm}+ (a_0 - \psi_0)\eta_b, 
 \vspace{3mm} \\ 
(c) \ \ \ 
\left(\bar{\nabla}_{\frac{\partial}{\partial x^b}}\eta^{\star}\right)\left(\frac{\partial}{\partial x^4}\right) 
= \Psi^2\delta_b^{\gamma}\psi_{\gamma} - \psi_0\Psi^2\eta_b\Phi^{-2} + \psi_4\eta_b. \end{array}\eqno(6.6)$$
Finally, taking into account (6.3)-(6.6), we deduce that

$$\begin{array}{l}
\bar{\nabla}_b\xi_a = - \Phi^{-2}\xi_b{\stackrel{.}{\xi}}_a 
+ \delta_a^{\alpha}\delta_b^{\beta}\left(\Theta_{\alpha\beta} + 
\Phi^2\omega_{\alpha\beta}\right) + \xi_a\delta_b^{\gamma}\phi_{\gamma}\vspace{2mm}\\ \hspace*{12mm}
+ \phi_4\Psi^{-2}\xi_a\eta_b   + 
\Psi^{-2}(\psi_0 - a_0)\eta_a\eta_b + \frac{1}{2}a_{\gamma}(\eta_a\delta^{\gamma}_b + \eta_b\delta^{\gamma}_a) 
\vspace{2mm}\\ \hspace*{12mm} +  \frac{1}{2}\Phi^2d_{\gamma}\Psi^{-2}(\eta_a\delta^{\gamma}_b 
- \eta_b\delta^{\gamma}_a),\end{array}\eqno(6.7)$$
and
$$\begin{array}{l}
\bar{\nabla}_b\eta_a = \Psi^{-2}\eta_b{\stackrel{.}{\eta}}_a + \delta_a^{\alpha}\delta_b^{\beta}\left(K_{\alpha\beta} 
-  \Psi^2\eta_{\alpha\beta}\right) + \eta_a\delta_b^{\gamma}\psi_{\gamma} - \psi_0\Phi^{-2}\eta_a\xi_b \vspace{2mm}\\ 
\hspace*{8mm} -
\phi_4\Phi^{-2}\xi_a\xi_b + \frac{1}{2}d_{\gamma}(\xi_a\delta^{\gamma}_b + \xi_b\delta^{\gamma}_a) 
+  \frac{1}{2}\Psi^2a_{\gamma}\Phi^{-2}(\xi_a\delta^{\gamma}_b 
- \xi_b\delta^{\gamma}_a).\end{array}\eqno(6.8)$$
Next, we note that each spatial tensor field defines a tensor field of the same type on $\bar{M}$. Here, we give 
some examples:

$$\begin{array}{l}\bar{\omega}_{ab} = \delta_a^{\alpha}\delta_b^{\beta}\omega_{\alpha\beta}; \ \ \ 
\bar{\Theta}_{ab} = \delta_a^{\alpha}\delta_b^{\beta}\Theta_{\alpha\beta}; \ \ \
\bar{h}_{ab} = \delta_a^{\alpha}\delta_b^{\beta}h_{\alpha\beta}; \\  
\bar{\sigma}_{ab} = \delta_a^{\alpha}\delta_b^{\beta}\sigma_{\alpha\beta} = \bar{\Theta}_{ab} 
- \frac{1}{3}\Theta\bar{h}_{ab}; \ \ \ 
\bar{\phi}_{a} = \delta_a^{\alpha}\phi_{\alpha}; \ \ \mbox{etc.}\end{array}$$
Taking into account this transformation process of spatial tensor fields into tensor fields, we express (6.7) 
and (6.8) as follows:

$$\begin{array}{l}
\bar{\nabla}_b\xi_a = - \Phi^{-2}\xi_b{\stackrel{.}{\xi}}_a + \bar{\sigma}_{ab} + \frac{1}{3}\Theta\bar{h}_{ab} + 
\Phi^2\bar{\omega}_{ab} + \xi_a\bar{\phi}_{b} + \phi_4\Psi^{-2}\xi_a\eta_b \vspace{2mm}\\ 
\hspace*{8mm} + 
\Psi^{-2}(\psi_0 - a_0)\eta_a\eta_b + \frac{1}{2}(\eta_a\bar{a}_b + \eta_b\bar{a}_a) 
 +  \frac{1}{2}\Phi^2\Psi^{-2}(\eta_a\bar{d}_b 
- \eta_b\bar{d}_a),\end{array}\eqno(6.9)$$
and
$$\begin{array}{l}
\bar{\nabla}_b\eta_a = \Psi^{-2}\eta_b{\stackrel{.}{\eta}}_a + \bar{H}_{ab} + \frac{1}{3}K\bar{h}_{ab} 
-  \Psi^2\bar{\eta}_{ab} + \eta_a\bar{\psi}_b - \psi_0\Phi^{-2}\eta_a\xi_b \vspace{2mm}\\ 
\hspace*{12mm} -
\phi_4\Phi^{-2}\xi_a\xi_b + \frac{1}{2}(\xi_a\bar{d}_b + \xi_b\bar{d}_a) 
 +  \frac{1}{2}\Psi^2\Phi^{-2}(\xi_a\bar{a}_b - \xi_b\bar{a}_a).\end{array}\eqno(6.10)$$
Now, we consider some particular cases. First, suppose that both $\xi$ and $\eta$ are unit vector fields, that is 
$\Phi^2 = \Psi^2 = 1$. Then, (6.9) and (6.10) become 

$$\begin{array}{l}
\bar{\nabla}_b\xi_a = - \xi_b{\stackrel{.}{\xi}}_a + \bar{\sigma}_{ab} + \frac{1}{3}\Theta \bar{h}_{ab} 
+ \bar{\omega}_{ab} - a_0\eta_a\eta_b \vspace{2mm} \\ \hspace*{18mm}+ \frac{1}{2}\{\eta_a(\bar{a}_b + \bar{d}_b) + 
\eta_b(\bar{a}_a - \bar{d}_a)\}, \end{array}\eqno(6.11)$$
and
$$\begin{array}{l}
\bar{\nabla}_b\eta_a = \eta_b{\stackrel{.}{\eta}}_a + \bar{H}_{ab} + \frac{1}{3}K\bar{h}_{ab} 
- \bar{\eta}_{ab} \vspace{2mm}\\ 
\hspace*{12mm} + \frac{1}{2}\{\xi_a(\bar{d}_b + \bar{a}_b ) +  \xi_b(\bar{d}_a - \bar{a}_a)\}.\end{array}
\eqno(6.12)$$\vspace{3mm}\newline
{\bf Remark 6.1} Note that the first four terms in the right hand side of (6.11) look formally as the ones 
in the (1+3) threading of a $4D$ spacetime (\cite{ej}, p.85). The other terms show the contribution of the 
fifth dimension in the kinematic theory of the $5D$ universe $(\bar{M}, \bar{g})$. $\blacksquare$ \newline
Finally, suppose that the following conditions are satisfied:\newline
(i) The distributions ${\cal{S}}\bar{M}\oplus{\cal{T}}\bar{M}$ and ${\cal{S}}\bar{M}\oplus{\cal{V}}\bar{M}$ 
are integrable.\newline
(ii) Both vector fields $\xi$ and $\eta$ define congruences of geodesics.\newline
By (4.4) we see that condition (i) is equivalent to (4.6) and 

$$a_{\alpha} = d_{\alpha} = 0, \ \ \forall \ \alpha \in \{1, 2, 3\}.\eqno(6.13)$$
Also, from (5.12h), (5.12i) and (6.4), we deduce that condition (ii) is equivalent to 

$$\begin{array}{l}(a) \ \ \ \phi_{\alpha} = b_{\alpha}, \ \ \ \phi_0 = 0, \ \ \ \phi_4 = 0, \ \ \ 
\stackrel{.}{\xi}_a = 0,\vspace{2mm} \\ 
(b) \ \ \ 
 \psi_{\alpha} = c_{\alpha}, \ \ \ \psi_0 = a_0, \ \ \ \psi_4 = 0, \ \ \ 
\stackrel{.}{\eta}_a = 0.\end{array} \eqno(6.14)$$
Then, by using (4.6), (6.13) and (6.14) into(6.9) and (6.10), we obtain

$$\bar{\nabla}_b\xi_a = \bar{\sigma}_{ab} + \frac{1}{3}\Theta\bar{h}_{ab} + \xi_a\bar{\phi}_{b},\eqno(6.15)$$
and
$$\bar{\nabla}_b\eta_a = \bar{H}_{ab} + \frac{1}{3}K\bar{h}_{ab} 
 + \eta_a\bar{\psi}_b - \psi_0\Phi^{-2}\eta_a\xi_b .\eqno(6.16)$$
It is an interesting (and difficult as well) question, to find solutions for Einstein equations in a $5D$ 
universe satisfying the conditions (i) and (ii).

\section{Raychaudhuri Equations in a $5D$ Universe}

As it is well known, the evolution of the expansion in a $4D$ spacetime is governed by Raychaudhuri equation, 
which also plays an important role in the proof of Penrose-Hawking singularity theorems. So, it is a need for a 
study of the evolutions of both the $4D$ and $5D$ expansions $\Theta$ and $K$, given by (4.11). Such a study 
leads us to some equations of Raychaudhuri type, expressing the derivatives of $\Theta$ and $K$ with respect 
to both variables $x^0$ (time) and $x^4$ (fifth dimension).\par
In what it follows, $\bar{R}$ denotes both the curvature tensor fields of the $5D$ universe $(\bar{M}, \bar{g})$ 
of type (0,4) and (1,3), given by
 
$$\begin{array}{l} \vspace{2mm}\\ 
(a) \ \ \ \bar{R}(X, Y, Z, U) = g(\bar{R}(X, Y, U), Z),\vspace{2mm}\\ 
(b) \ \ \ \bar{R}(X, Y, U) = \bar{\nabla}_X\bar{\nabla}_YU - \bar{\nabla}_Y\bar{\nabla}_XU - 
\bar{\nabla}_{[X, Y]}U,\end{array}\eqno(7.1)$$
for all $X, Y, Z, U \in \Gamma(T\bar{M})$. Now, we consider an orthonormal frame field $\{E_k, 
\Phi^{-1}\frac{\delta }{\delta x^0}, \Psi^{-1}\frac{\partial }{\partial x^4}\}$, where $\{E_{\gamma}\}, \ 
\gamma \in \{1,2,3\}$, is an orthonormal basis in $\Gamma({\cal{S}}\bar{M})$. Then, we put

$$E_{\gamma} = E_{\gamma}^{\alpha}\frac{\delta }{\delta x^{\alpha}},\eqno(7.2)$$
and deduce that 

$$h^{\alpha\beta} =  \sum_{\gamma =1}^3E_{\gamma}^{\alpha}E_{\gamma}^{\beta}.\eqno(7.3)$$
By evaluating some local components of the Ricci tensor $\bar{Ric}$ of $(\bar{M}, \bar{g})$, we shall obtain equations 
of Raychaudhuri type about $\Theta$ and $K$. According to (\cite{on}, p.87), and using (7.2) and (7.3), we obtain
$$\begin{array}{c}
\bar{Ric}(X, Y) = h^{\alpha\beta}\bar{R}(\frac{\delta}{\delta x^{\beta}}, X, \frac{\delta}{\delta x^{\alpha}}, Y) 
- \Phi^{-2}\bar{R}(\frac{\delta }{\delta x^0}, X, \frac{\delta }{\delta x^0}, Y),\vspace{2mm}\\
+\Psi^{-2}\bar{R}(\frac{\partial }{\partial x^4}, X, \frac{\partial }{\partial x^4}, Y), \ \forall \ X, Y 
\in \Gamma(T\bar{M}).\end{array}$$
For our purpose, we consider only the local components:

$$\begin{array}{l}(a) \ \ \ \bar{R}_{00} = \bar{Ric}(\frac{\delta }{\delta x^0}, \frac{\delta }{\delta 
 x^0}) = h^{\alpha\beta}\bar{R}(\frac{\delta }{\delta x^{\beta}}, \frac{\delta }{\delta 
 x^0}, \frac{\delta }{\delta x^{\alpha}}, \frac{\delta }{\delta x^0})\vspace{2mm}\\ \hspace*{48mm} + \Psi^{-2}
\bar{R}(\frac{\partial }{\partial x^4}, \frac{\delta }{\delta x^0}, \frac{\partial }{\partial x^4}, 
 \frac{\delta }{\delta x^0}), \vspace{3mm}\\
 (b) \ \ \ \bar{R}_{44} = \bar{Ric}(\frac{\partial }{\partial  x^4}, \frac{\partial }{\partial 
 x^4}) = h^{\alpha\beta}\bar{R}(\frac{\delta }{\delta x^{\beta}}, \frac{\partial }{\partial 
 x^4}, \frac{\delta }{\delta x^{\alpha}}, \frac{\partial }{\partial x^4})\vspace{2mm}\\ \hspace*{48mm} 
- \Phi^{-2}\bar{R}(\frac{\delta }{\delta x^0}, \frac{\partial }{\partial x^4}, \frac{\delta }{\delta x^0}, 
\frac{\partial }{\partial x^4}), \vspace{3mm}\\ 
(c) \ \ \ \bar{R}_{04} = \bar{Ric}(\frac{\delta }{\delta x^0}, \frac{\partial }{\partial 
 x^4}) = h^{\alpha\beta}\bar{R}(\frac{\delta }{\delta x^{\beta}}, \frac{\delta }{\delta 
 x^0}, \frac{\delta }{\delta x^{\alpha}}, \frac{\partial }{\partial x^4})
.\end{array}\eqno(7.4)$$
Next, we show in detail the calculations for the right hand side of (7.4a). First, by using the symmetries of 
$\bar{R}$ and (7.1a), we obtain

$$\begin{array}{l}
\bar{R}(\frac{\delta }{\delta x^{\beta}}, \frac{\delta }{\delta 
 x^0}, \frac{\delta }{\delta x^{\alpha}}, \frac{\delta }{\delta x^0})  = 
\bar{R}(\frac{\delta }{\delta x^0}, \frac{\delta }{\delta x^{\beta}}, \frac{\delta }{\delta x^0}, 
\frac{\delta }{\delta x^{\alpha}})\vspace{2mm}\\ \hspace*{36mm} = \bar{g}(\bar{R}(\frac{\delta }{\delta x^0}, 
 \frac{\delta }{\delta x^{\beta}}, \frac{\delta }{\delta x^{\alpha}}), \frac{\delta }{\delta x^0}).
\end{array}\eqno(7.5)$$
Then, denote by ${\cal{T}}$ the projection morphism of $\Gamma(T\bar{M})$ to $\Gamma({\cal{T}}\bar{M})$ with 
respect to (2.12), and by using (5.12), (5.13a) and (4.4a), we infer that

$$\begin{array}{lc}(a) \ \ \ {\cal{T}}\bar{\nabla}_{\frac{\delta }{\delta x^0}}\bar{\nabla}_{\frac{\delta }
{\delta x^{\beta}}}\frac{\delta }{\delta x^{\alpha}} = \left\{\frac{\delta\omega_{\alpha\beta}}{\delta x^0} + 
\Phi^{-2}\frac{\delta\Theta_{\alpha\beta}}{\delta x^0} + (\phi_{\gamma} - b_{\gamma})\Gamma^{\ \;\gamma}_{\alpha 
\ \;\beta}\right.\vspace{2mm}\\ \left.\hspace*{18mm} + \phi_0(\omega_{\alpha\beta} - \Phi^{-2}\Theta_{\alpha\beta}) 
+  \phi_4(\eta_{\alpha\beta} - \Psi^{-2}K_{\alpha\beta})\right\}\frac{\delta }{\delta x^0},\vspace{2mm}\\   
(b) \ \ \ {\cal{T}}\bar{\nabla}_{\frac{\delta }{\delta x^{\beta}}}\bar{\nabla}_{\frac{\delta }{\delta x^0}}
\frac{\delta }{\delta x^{\alpha}} = \left\{\frac{\delta }{\delta x^{\beta}}(\phi_{\alpha} - b_{\alpha}) + 
(\phi_{\alpha} - b_{\alpha})\phi_{\beta} + (\omega_{\gamma\beta}\right.\vspace{2mm}\\ \left.\hspace*{6mm} 
+ \Phi^{-2}\Theta_{\gamma\beta})\Gamma^{\ \;\gamma}_{\alpha \ \;0} + \frac{1}{4}(\Phi^2d_{\alpha}\Psi^{-2} 
- a_{\alpha})(\Psi^{2}a_{\beta}\Phi^{-2} + d_{\beta})\right\}\frac{\delta }{\delta x^0},\vspace{2mm}\\ 
(c) \ \ \ {\cal{T}}\bar{\nabla}_{\left[\frac{\delta }{\delta x^0}, \frac{\delta }{\delta x^{\beta}}\right]}
\frac{\delta }{\delta x^{\alpha}} = \left\{(b_{\alpha} - \phi_{\alpha})b_{\beta} +  \frac{1}{2}(d_{\alpha} - 
\Psi^2a_{\alpha}\Phi^{-2})a_{\beta}\right \}\frac{\delta }{\delta x^0}.\end{array}\eqno(7.6)$$
Taking into account (7.1b) and (7.6), and using covariant derivatives induced by the Riemannian spatial 
connection (see (5.10)), we deduce that

$$\begin{array}{lc}{\cal{T}}\bar{R}(\frac{\delta }{\delta x^0}, \frac{\delta }{\delta x^{\beta}}, 
\frac{\delta }{\delta x^{\alpha}}) = \left\{\omega_{\alpha\beta|_0} + \Phi^{-2}\Theta_{\alpha\beta|_0} + 
(\omega_{\alpha\gamma} + \Phi^{-2}\Theta_{\alpha\gamma})\Gamma^{\ \;\gamma}_{\beta \ \;0}\right.\vspace{2mm}\\ 
\left. \hspace*{14mm}+  \phi_0(\omega_{\alpha\beta} - \Phi^{-2}\Theta_{\alpha\beta}) 
+  \phi_4(\eta_{\alpha\beta} - \Psi^{-2}K_{\alpha\beta}) - (\phi_{\alpha} - b_{\alpha})_{|_{\beta}} 
 \right.\vspace{2mm}\\ \left.\hspace*{14mm} - (\phi_{\alpha} - b_{\alpha})(\phi_{\beta} - b_{\beta}) 
 + \frac{1}{4}( a_{\alpha} - \Phi^2d_{\alpha}\Psi^{-2})d_{\beta}\right.\vspace{2mm}\\ 
\left. \hspace*{14mm} - \frac{3}{4}(d_{\alpha} - \Psi^{2}a_{\alpha}\Phi^{-2})a_{\beta}\right\}
\frac{\delta }{\delta x^0}.\end{array}\eqno(7.7)$$
Now, we put 
 
$$\begin{array}{c}(a) \ \ \omega_{\alpha\beta}\omega^{\alpha\beta} = \omega^2, \ \ \ (b) \ \ 
\ \sigma_{\alpha\beta}\sigma^{\alpha\beta} = \sigma^2,\vspace{2mm}\\ \ \ (c) \ \ \ \eta_{\alpha\beta}
\eta^{\alpha\beta} = \eta^2, \ \ \ (d) \ \ \ H_{\alpha\beta}H^{\alpha\beta} = H^2,\end{array}\eqno(7.8) $$
and obtain 

$$\begin{array}{lc}(a) \ \ \ \omega_{\gamma}^{\beta}\omega^{\gamma}_{\beta} = -\omega^2, \ \ \ (b) \ \ 
\ \Theta_{\gamma}^{\beta}\Theta^{\gamma}_{\beta} = \sigma^2 + \frac{1}{3}\Theta^2,\vspace{2mm}\\ \ \ (c) 
\ \ \ \eta_{\gamma}^{\beta}\eta^{\gamma}_{\beta} = - \eta^2, \ \ \ (d) \ \ \ K_{\gamma}^{\beta}K^{\gamma}_{\beta} 
= H^2 + \frac{1}{3}K^2.\end{array}\eqno(7.9) $$
Moreover, by using the symmetries of expansion and vorticity tensor fields, we infer that 

$$(a) \ \ \ \Theta_{\gamma}^{\beta}\omega^{\gamma}_{\beta} = 0,  \ \ \ (b) \ \ 
\ K_{\gamma}^{\beta}\eta^{\gamma}_{\beta} = 0.\eqno(7.10) $$
Contracting (7.5) by $h^{\alpha\beta}$ and using (7.7), (2.11), (5.9a), (5.11b), (7.9a), (7.9b) and (7.10a), 
we obtain

$$\begin{array}{l} h^{\alpha\beta}\bar{R}(\frac{\delta }{\delta x^{\beta}}, \frac{\delta }{\delta 
 x^0}, \frac{\delta }{\delta x^{\alpha}}, \frac{\delta }{\delta x^0})  = -\Theta_{|_0} - \sigma^2 -
\frac{1}{3}\Theta^2 - \frac{3}{4}\Psi^2a^2 + \phi_0\Theta \vspace{2mm}\\ 
+ \Phi^2\left\{\phi_4\Psi^{-2}K + (\phi^{\gamma} - b^{\gamma})_{|_{\gamma}} + (\phi^{\gamma} 
- b^{\gamma})(\phi_{\gamma} - b_{\gamma}) + \frac{1}{2}a_{\gamma}d^{\gamma}\right\}\vspace{2mm}\\ 
+ \Phi^4\left\{\omega^2 + \frac{1}{4}\Psi^{-2}d^2\right\}, \end{array} \eqno(7.11)$$ 
 where we put 
$$\begin{array}{l} 
\Theta_{|_0} = \frac{\delta \Theta}{\delta x^0}, \ \ \ \ a^2 = a_{\gamma}a^{\gamma}, \ \ \ \ d^2 = d_{\gamma}d^{\gamma}
.\end{array}$$  
Similar calculations as performed for (7.7) lead us to the following:
 
$$\begin{array}{lc}{\cal{T}}\bar{R}(\frac{\delta }{\delta x^0}, \frac{\partial }{\partial x^4}, 
\frac{\partial }{\partial x^4}) = \left\{\Psi^2(c^{\gamma} - \psi^{\gamma})(\phi_{\gamma} - b_{\gamma})
-\frac{1}{4}\Psi^{4}a^2\Phi^{-2}\right.\vspace{2mm}\\ \left. - \frac{1}{4}\Phi^2d^2 - (\phi_4)^2
- \frac{\partial \phi_4}{\partial x^4} + \frac{1}{2}\Psi^2a_{\gamma}d^{\gamma} + \phi_{4}\psi_4 + 
[\Psi^2(\psi_0 - a_0)\Phi^{-2}]_{|_0}\right.\vspace{2mm}\\ \left. + \Psi^2(\psi_0 - a_0)(\phi_0 -\psi_0 
- a_0)\Phi^{-2}\right\}\frac{\delta }{\delta x^0}.\end{array}\eqno(7.12)$$
Then, by using (7.1a), (7.12) and (2.11), we deduce that 

$$\begin{array}{r}\bar{R}(\frac{\partial }{\partial x^4}, \frac{\delta }{\delta x^0}, 
\frac{\partial }{\partial x^4},\frac{ \delta }{\delta x^0}) = \Phi^2\Psi^2\{(\phi^{\gamma} - b^{\gamma})
(\psi_{\gamma} - 
c_{\gamma}) - \frac{1}{2}a_{\gamma}d^{\gamma}\}\vspace{2mm}\\  + \Phi^2\{(\phi_4)^2  
+ \frac{\partial \phi_4}{\partial x^4} - \phi_4\psi_4\}  + \Psi^2\left\{(\psi_0 - a_0)(\phi_0 -\psi_0 + a_0) 
\right.\vspace{2mm}\\ \left. - (\psi_0 - a_0)_{|_0}\right.\} + \frac{1}{4}\{\Phi^4d^2 + \Psi^4a^2\}. 
\end{array}\eqno(7.13)$$
Finally, by using (7.11) and (7.13) into (7.4a), we infer that

$$\begin{array}{lc}\Theta_{|_0} = -\sigma^2 - \frac{1}{3}\Theta^2 - \frac{1}{2}\Psi^2a^2 + \phi_0\Theta - 
(\psi_0 - a_0)_{|_0}\vspace{2mm}\\ \hspace*{10mm} + (\psi_0 - a_0)(\phi_0 -\psi_0 + a_0) 
+ \Phi^2\left\{\phi_4\Psi^{-2}K + (\phi^{\gamma} - b^{\gamma})_{|_{\gamma}} 
\right.\vspace{2mm}\\ \left. \hspace*{10mm} + (\phi_{\gamma} - b_{\gamma})(\phi^{\gamma} - b^{\gamma} - c^{\gamma} 
+ \psi^{\gamma}) + \Phi^2(\omega^2 + \frac{1}{2}\Psi^{-2}d^2)\right.\vspace{2mm}\\ \left.\hspace*{10mm} + \Psi^{-2}
\left((\phi_4)^2 + \frac{\partial \phi_4}{\partial x^4} - \phi_4\psi_4\right)\right\} - \bar{R}_{00}.
\end{array}\eqno(7.14)$$
We call (7.14) the $4D$ {\it Raychaudhuri equation} for the $4D$ expansion $\Theta$ of the $5D$ universe 
$(\bar{M}, \bar{g}).$\par
By similar calculations, we obtain

$$\begin{array}{l}
\bar{R}(\frac{\delta }{\delta x^{\beta}}, \frac{\partial }{\partial x^4}, \frac{\delta }{\delta x^{\alpha}},
 \frac{\partial }{\partial x^4}) = \bar{g}(\bar{R}(\frac{\partial }{\partial x^4}, 
 \frac{\delta }{\delta x^{\beta}}, \frac{\delta }{\delta x^{\alpha}}), \frac{\partial }{\partial x^4})\vspace{2mm}\\ 
= \Psi^2\left\{\eta_{\alpha\beta|4} - \Psi^{-2}K_{\alpha\beta|4} - (\psi_{\alpha} - c_{\alpha})_{|\beta} + 
 (\eta_{\alpha\gamma} - \Psi^{-2}K_{\alpha\gamma})\Gamma^{\ \gamma}_{\beta \;4}\right.\vspace{2mm}\\ \left. 
+ (\omega_{\alpha\beta} + \Phi^{-2}\Theta_{\alpha\beta})(\psi_0 - a_0) + (\eta_{\alpha\beta} 
+ \Psi^{-2}K_{\alpha\beta})\psi_4 + \frac{3}{4}(\Phi^2d_{\alpha}\Psi^{-2}\right.\vspace{2mm}\\ \left. 
- a_{\alpha})d_{\beta} + \frac{1}{4}(d_{\alpha} - \Psi^2a_{\alpha}\Phi^{-2})a_{\beta} - (\psi_{\alpha} 
- c_{\alpha})(\psi_{\beta} - c_{\beta})\right\}.\end{array}\eqno(7.15)$$
Then, by using (7.15) and (7.13) into (7.4b), we deduce that

$$\begin{array}{lc} K_{|4} = -H^2 - \frac{1}{3}K^2 + \Psi^4\eta^2 - \Psi^2\left\{(\psi^{\gamma} - 
c^{\gamma})_{|\gamma} + (\psi^{\gamma} - c^{\gamma})(\phi_{\gamma}\right.\vspace{2mm}\\ \left. - b_{\gamma} 
+ \psi_{\gamma} - c_{\gamma})\right\} + \Psi^2\Phi^{-2}\left\{(\psi_0 - a_0)_{|0} + (\psi_0 - a_0)(\psi_0 - 
\phi_0 \right. \vspace{2mm}\\ \left. - a_0 + \Theta) - \frac{1}{2}\Psi^2a^2\right\} + \phi_4K + 
\frac{1}{2}\Phi^2d^2 - (\phi_4)^2 - \frac{\partial\phi_4}{\partial x^4}\vspace{2mm}\\  
+ \phi_4\psi_4 - \bar{R}_{44},\end{array}\eqno(7.16)$$
where we put

$$K_{|4} = \frac{\partial K}{\partial x^4}.$$
We call (7.16) the $5D$ {\it Raychaudhuri equation} for the $5D$ expansion $K$ of the $5D$ universe 
$(\bar{M}, \bar{g}).$\par
Finally, we shall state some Raychaudhuri equations which involve 

$$\Theta_{|4} = \frac{\partial \Theta}{\partial x^4}, \ \ \ \mbox{and} \ \ \ K_{|0} = \frac{\delta K}{\delta x^0}.$$
First, by using (7.1), (5.12), (4.4a) and (2.3), we infer that 

 $$\begin{array}{l}
\bar{R}(\frac{\delta }{\delta x^{\beta}}, \frac{\delta }{\delta 
 x^0}, \frac{\delta }{\delta x^{\alpha}}, \frac{\partial }{\partial x^4}) = \bar{R}(\frac{\delta }{\delta  x^0}, 
 \frac{\delta }{\delta x^{\beta}}, \frac{\partial }{\partial x^4},  \frac{\delta }{\delta x^{\alpha}}) 
 \vspace{2mm}\\ = \bar{g}(\bar{R}(\frac{\delta }{\delta x^0}, \frac{\delta }{\delta x^{\beta}}, 
\frac{\delta }{\delta x^{\alpha}}), \frac{\partial }{\partial x^4})  
= \Psi^2\eta_{\alpha\beta|0} - K_{\alpha\beta|0} + \frac{1}{2}\left(a_{\alpha} \right. \vspace{2mm}\\ \left.
 - \Phi^2d_{\alpha}\Psi^{-2}\right)_{|\beta}\Psi^2 + (\Psi^2\eta_{\alpha\gamma} - K_{\alpha\gamma})
\Gamma^{\ \gamma}_{\beta \;0} + \Psi^2(\psi_{\alpha} - c_{\alpha})a_{\beta}\vspace{2mm}\\  + 
(\Psi^2\eta_{\alpha\beta} + K_{\alpha\beta})\psi_0 - \frac{1}{2}(\phi_{\alpha} - b_{\alpha})(\Phi^2d_{\beta} 
+ \Psi^2a_{\beta})\vspace{2mm}\\ - \frac{1}{2}(\Phi^2d_{\alpha} - \Psi^2a_{\alpha})(\psi_{\beta} - b_{\beta}) + 
(\Phi^2\omega_{\alpha\beta} + \theta_{\alpha\beta})\phi_4.\end{array}\eqno(7.17)$$
Contracting (7.17) by $h^{\alpha\beta}$, and using (7.4c), (4.11), (5.9a) and (5.11b), we obtain 

$$\begin{array}{l}K_{|0} = \frac{1}{2}(a^{\gamma} - \Phi^2d^{\gamma}\Psi^{-2})_{|\gamma}\Psi^2 
- K_{\alpha\beta}\Theta^{\alpha\beta} + \psi_0K + \phi_4\Theta \vspace{2mm}\\ \hspace*{10mm} - 
 \frac{1}{2}\Phi^2d_{\gamma}(\phi^{\gamma} + \psi^{\gamma} - 2b^{\gamma})
+ \Psi^2\left\{\frac{1}{2}a_{\gamma}(3\psi^{\gamma} - \phi^{\gamma} - 2c^{\gamma}) \right.\vspace{2mm}\\ 
\hspace*{10mm}\left.
- \Phi^2\eta_{\alpha\beta}\omega^{\alpha\beta}\right\} - \bar{R}_{04}.\end{array}\eqno(7.18)$$
On the other hand, we calculate $\bar{R}$ from (7.4c) by using (7.1), (5.12), (4.4b) and (2.11), and deduce that 

$$\begin{array}{l}
\bar{R}(\frac{\delta }{\delta x^{\beta}}, \frac{\delta }{\delta 
 x^0}, \frac{\delta }{\delta x^{\alpha}}, \frac{\partial }{\partial x^4}) = \bar{R}(\frac{\partial }{\partial x^4}, 
  \frac{\delta }{\delta x^{\alpha}}, \frac{\delta }{\delta  x^0}, \frac{\delta }{\delta x^{\beta}}) 
 \vspace{2mm}\\ = \bar{g}(\bar{R}(\frac{\partial }{\partial x^4}, \frac{\delta }{\delta x^{\alpha}}, 
\frac{\delta }{\delta x^{\beta}}),\frac{\delta }{\delta x^0}) = - \Theta_{\alpha\beta|4} + 
\Phi^2\omega_{\alpha\beta|4} - \frac{1}{2}(d_{\beta} \vspace{2mm}\\  - \Psi^2a_{\beta}\Phi^{-2})_{|\alpha}\Phi^2 
+ (\Phi^2\omega_{\gamma\beta} - \theta_{\gamma\beta})\Gamma^{\ \gamma}_{\alpha \:4} + (\Phi^2\omega_{\alpha\beta} 
 +\theta_{\alpha\beta})\phi_4\vspace{2mm}\\ + \Psi^2(\psi_0 - a_0)(\eta_{\alpha\beta} + \Psi^{-2}K_{\alpha\beta}) + 
\frac{1}{2}\phi_{\alpha}(\Psi^2a_{\beta} - \Phi^2d_{\beta}) + \frac{1}{2}(\Psi^2a_{\alpha}\vspace{2mm}\\ 
+ \Phi^2d_{\alpha})(\psi_{\beta} - c_{\beta}) + \Phi^2d_{\alpha}(b_{\beta} - \phi_{\beta}) 
+ \frac{1}{2}c_{\alpha}(\Phi^2d_{\beta} - \Psi^2a_{\beta}).\end{array}\eqno(7.19)$$
Then, by using (7.19) into (7.4c), and taking into account (4.11), (5.11b) and (5.9b), we infer that

$$\begin{array}{l}\Theta_{|4} = \frac{1}{2}(\Psi^2a^{\gamma}\Phi^{-2} - d^{\gamma})_{|\gamma}\Phi^2  
- K_{\alpha\beta}\Theta^{\alpha\beta} + (\psi_0 - a_0)K + \phi_4\Theta \vspace{2mm}\\ \hspace*{18mm} 
- \frac{1}{2}\Phi^2d_{\gamma}(3\phi^{\gamma} -\psi^{\gamma} 
- 2b^{\gamma})  + \Psi^2\left\{\frac{1}{2}a_{\gamma}(\phi^{\gamma} + \psi^{\gamma} - 2c^{\gamma}) 
\right.\vspace{2mm}\\ \left. \hspace*{18mm} - \Phi^2\eta_{\alpha\beta}\omega^{\alpha\beta}\right\}
- \bar{R}_{04}.\end{array}\eqno(7.20)$$
By elementary calculations using (4.7), we obtain 

$$\begin{array}{l}(a) \ (\Psi^2a^{\gamma}\Phi^{-2} - d^{\gamma})_{|\gamma}\Phi^2=(\Psi^2a^{\gamma} 
- \Phi^2d^{\gamma})_{|\gamma}\vspace{2mm}\\ \hspace*{44mm} + 2\phi_{\gamma}(\Phi^2d^{\gamma} - 
\Psi^2a^{\gamma}), \vspace{2mm}\\ 
(b) \ (a^{\gamma} - \Phi^2d^{\gamma}\Psi^{-2})_{|\gamma}\Psi^2=(\Psi^2a^{\gamma} 
- \Phi^2d^{\gamma})_{|\gamma}\vspace{2mm}\\ \hspace*{44mm} + 2\psi_{\gamma}(\Phi^2d^{\gamma} - 
\Psi^2a^{\gamma}).\end{array}\eqno(7.21)$$
Finally, by using (7.21a) into (7.20), and (7.21b) into (7.18), we deduce that

$$\begin{array}{l}\Theta_{|4} = K_{|0} - a_0K = \frac{1}{2}(\Psi^2a^{\gamma} - \Phi^2d^{\gamma})_{|\gamma} 
- K_{\alpha\beta}\Theta^{\alpha\beta} - \Phi^2\Psi^2\eta_{\alpha\beta}\omega^{\alpha\beta}\vspace{2mm}\\ 
\hspace*{28mm} + (\psi_0 - a_0)K + \phi_4\Theta + \frac{1}{2}\Phi^2(\Psi^{\gamma} -\phi^{\gamma} 
+ 2b^{\gamma})d_{\gamma}
\vspace{2mm}\\ \hspace*{28mm} + \frac{1}{2}\Psi^2(\psi^{\gamma} -\phi^{\gamma} - 2c^{\gamma})a_{\gamma} 
- \bar{R}_{04}.\end{array}\eqno(7.22)$$
We call (7.22) the {\it mixed Raychaudhuri equation} in the $5D$ universe $(\bar{M}, \bar{g})$.\par

 The classical  Raychaudhuri equation in a $4D$ spacetime has ben generalized in several directions. We recall 
here two of such generalizations. First, in \cite{av} there were stated Raychaudhuri equations for single and 
two non-normalized vector fields in a $4D$ spacetime. Our Raychaudhuri equations are also with respect to two 
 non-normalized vector fields, but in a $5D$ universe. Then, in \cite{ca} there have been obtained generalizations 
of Raychaudhuri equation for the evolution of deformations of a relativistic membrane of arbitrary dimension in 
an arbitrary background spacetime. The approach we developed in this paper is totally different from the one  
presented in \cite{ca}.\par
Note that the above Raychaudhuri equations have been obtained in the most general $5D$ universe. In particular, 
 suppose that the lift of $\partial /\partial x^0$ from the $4D$ spacetime to the $5D$ universe is orthogonal to 
$\partial /\partial x^4$. Then, by (2.9b) and (2.10) we deduce that $A_0 =0$, which implies 
 
 $$\delta /\delta x^0 = \partial /\partial x^0, \eqno(7.23)$$
 via (2.7). Now, each point P of $\bar{M}$ can be considered as intersection point of an integral curve of 
$\partial /\partial x^0$ on which we choose $x^0$ as parameter, with an integral curve of $\partial /\partial x^4$ 
parametrized by $x^4$. Since in this case $a_0 = 0$ (see (3.1b)), from (7.22) and (7.23) we deduce that

$$\frac{d\Theta}{dx^4} = \frac{dK}{dx^0},\eqno(7.24)$$
 at any point $P\in\bar{M}$. Thus, {\it the rate of change of the $4D$ expansion $\Theta$ in the direction of 
the fifth dimension, is equal to the rate of change of the $5D$ expansion $K$ in the time direction}.\par
Finally, we prove the following important results for the kinematic theory of a $5D$ universe.\newline 
 
{\bf Theorem 7.1.} {\it Let $(\bar{M}, \bar{g})$ be a $5D$ universe satisfying the following conditions}:\newline
(a) {\it The distribution ${\cal{S}}\bar{M}\oplus {\cal{V}}\bar{M}$ is integrable}.\newline
(b) {\it $\xi = \delta/\delta x^0$ is given by} (7.23), {\it and define a congruence of geodesics}.\newline
(c) $\bar{R}_{00}\geq 0$, {\it and $\psi_{0|0} \geq 0$ on $\bar{M}$}.\newline
{\it If the $4D$ expansion $\Theta$ takes the negative value $\Theta_0$ at a point of a geodesic in the congruence,
then $\Theta$ goes to $-\infty$ along that geodesic and the proper time coordinate $x^0$ satisfies} 
$x^0 \leq -3/{\Theta_0}.$\vspace{3mm}\newline
{\bf Proof.} Taking into account (4.4b) and (4.4c), the condition (a) is equivalent to 

$$d_{\alpha} = 0 \ \ \ \mbox{and} \ \ \ \omega_{\alpha\beta} = 0, \ \ \forall \ \alpha, \beta \in\{1,2,3\}.
\eqno(7.25) $$
Then, note that condition (b) implies $A_0 = 0$ and (6.14a). Taking into account that $a_0 = 0$ (cf.(3.1b)), and 
using (7.25) and (6.14a) into the $4D$ Raychaudhuri equation (7.14), we obtain

$$\Theta_{|0} = -\sigma^2 - \frac{1}{3}\Theta^2 - \frac{1}{2}\Psi^2a^2 -(\psi_0)^2 - \psi_{0|0} -\bar{R}_{00}.
\eqno(7.26)$$
Finally, by condition (c) we see that (7.26) implies 

$$\Theta_{|0} + \frac{1}{3}\Theta^2 \leq 0,\eqno(7.27)$$
along any geodesic $C$ that is tangent to $\xi$. Choose $x^0$ as parameter on $C$ and by (7.23), we express (7.27) 
as follows

$$\frac{d\Theta}{dx^0} + \frac{1}{3}\Theta^2(x^0) \leq 0,\eqno(7.28)$$
along $C$. By using (7.28), and the same reason as in the kinematic theory of a $4D$ spacetime (see the proof 
of Lemma 9.2.1 in \cite{w}), we obtain the assertion of the theorem. $\blacksquare$ \vspace{3mm}\newline
{\bf Remark 7.1} The conditions (7.23) and $\psi_{0|0}\geq 0$ are new conditions comparing with a similar result 
in a $4D$ spacetime. We note that (7.23) is equivalent to $\bar{g}_{04} = 0$, which is satisfied by all the 
$5D$ models from both the brane-world theory \cite{mk} and the space-time-matter theory \cite{o}. Also, $\psi_{0|0} 
= 0$ for any $5D$ universe in which $\eta$ is a unit vector field. $\blacksquare$ \vspace{3mm}\newline
{\bf Theorem 7.2.} {\it Let $(\bar{M}, \bar{g})$ be a $5D$ universe satisfying the following conditions}:\newline
(a) {\it The distributions ${\cal{S}}\bar{M}\oplus {\cal{V}}\bar{M}$ and ${\cal{S}}\bar{M}\oplus {\cal{T}}\bar{M}$ 
are integrable}.\newline
(b) {\it Both vector fields $\xi$ and $\eta$ define congruences of geodesics}.\newline
(c) $\bar{R}_{44}\geq 0$, {\it on $\bar{M}$}.\newline
{\it If the $5D$ expansion $K$ takes the negative value $K_0$ at a point of a geodesic determined by $\eta$,
then $K$ goes to $-\infty$ along that geodesic and the fifth coordinate $x^4$ satisfies} 
$x^4 \leq -3/{K_0}.$\vspace{3mm}\newline
{\bf Proof.} The condition (a) is equivalent to (7.25) and 

$$a_{\alpha} = 0 \ \ \ \mbox{and} \ \ \ \eta_{\alpha\beta} = 0, \ \ \forall \ \alpha, \beta \in\{1,2,3\},
\eqno(7.29) $$
via (4.4a) and (4.4c). Also, note that condition (b) is equivalent to (6.14). Then, by using (7.25), (7.29) and 
(6.14) into the $5D$ Raychaudhuri equation (7.16), we obtain

$$K_{\vert_4} = -H^2 - \frac{1}{3}K^2 - \bar{R}_{44}.\eqno(7.30)$$ 
The condition (c) and (7.30) imply

$$K_{|4} + \frac{1}{3}K^2 \leq 0,\eqno(7.31)$$ 
on any geodesic $C$ defined by $\eta$. Consider $x^4$ as parameter on $C$ and by using (7.31) and the same 
reason as in Theorem 7.1, we obtain the assertion of the theorem.  $\blacksquare$ \vspace{3mm}\newline 
The Theorems 7.1 and 7.2 
might be useful in an attempt to prove singularity theorems for a $5D$ universe. Also, the three types 
 of Raychaudhuri equations (7.14), (7.16) and (7.22) might have an important role in a study of the evolution 
of a concrete $5D$ universe.

\section{Conclusions}
In the present paper, for the first time in literature, we develop a kinematic theory in a $5D$ universe. The 
main tools in our approach are the spatial tensor fields and the Riemannian connection defined on the spatial 
 distribution. It is worth mentioning that all the kinematic quantities (acceleration, expansion, shear and 
vorticity) are defined as spatial tensor fields, and therefore they should be considered as $3D$ geometric 
objects in a $5D$ universe.\par
Now, we stress on the novelty brought by our paper into the study of a $5D$ universe. First, we mention the 
(1+1+3) splitting determined by the two vector fields $\xi$ and $\eta$, which has an important role in relating 
 geometry and physics to the observations. In this way, the $5D$ universe is filled up by nets determined by 
integral curves of both $\xi$ and $\eta$. So, apart from the $4D$ kinematic quantities related to $\xi$, should 
be taken into consideration those determined by the assumption on the existence of the fifth dimension. Also, it 
is important to mention the three types of Raychaudhuri equations, which we obtained in the $5D$ universe $(\bar{M}, 
\bar{g})$. They describe the evolution of both the $4D$ expansion  and  $5D$ expansion along the two 
congruences determined by $\xi$ 
 and $\eta$. The Theorems 7.1 and 7.2 prove the existence of the singularities in both expansions $\Theta$ and $K$. 
Actually, these theorems state that caustics will develop in both congruences if convergence occurs anywhere. 
Taking into account the methods used to prove the Penrose-Hawking singularity theorems in a $4D$ spacetime, we 
think that such theorems might play an important role in a proof of the existence of singularities in a $5D$ 
universe.\par
Finally, we should mention that our study is mainly developed on the mathematical part of the kinematic theory 
in a $5D$ universe. This must be followed by detailed studies which might bring new insights on the $4D$ physics 
in the presence of the fifth dimension.

\end{document}